\documentclass[sigconf]{acmart}

\AtBeginDocument{%
  \providecommand\BibTeX{{%
    \normalfont B\kern-0.5em{\scshape i\kern-0.25em b}\kern-0.8em\TeX}}}

\usepackage[flushleft]{threeparttable}
\usepackage{bbm}
\usepackage{dblfloatfix} 
\usepackage{tabularx}
\usepackage{graphicx}
\usepackage{algorithmicx}
\usepackage[ruled,noend]{algorithm2e}
\usepackage{amsmath}
\usepackage{multirow}
\usepackage{tcolorbox}
\usepackage{booktabs}
\usepackage{multicol}
\usepackage{tabularx}
\usepackage{array}
\usepackage{colortbl}

\tcbuselibrary{skins}
\newcolumntype{Y}{>{\raggedleft\arraybackslash}X}

\setcopyright{acmcopyright}
\copyrightyear{2024}
\acmYear{2024}

\acmConference[***]{}{Oct 21--25,
  2024}{**}
\acmPrice{15.00}
\acmISBN{978-1-4503-XXXX-X/18/06}

\begin{document}

\title{Sequential Recommendation via Adaptive Robust Attention with Multi-dimensional Embeddings}

\author{Linsey Pang}
\email{xpang@salesforce.com}
\affiliation{%
\institution{Salesforce}
  \country{USA}
}
\author{Amir Hossein Raffiee}
\email{amir.raffiee@salesforce.com}
\affiliation{%
  \institution{Salesforce}
  \country{USA}
}
\author{Wei Liu}
\email{Wei.Liu@uts.edu.au}
\affiliation{
  \institution{University of Technology Sydney}
\country{Australia}
  }
\author{Keld Lundgaard}
\email{klundgaard@salesforce.com}
\affiliation{%
  \institution{Salesforce}
  \country{USA}
}

\begin{abstract}
Sequential recommendation models have achieved state-of-the-art performance using self-attention mechanism. It has since been found that moving beyond only using item ID and positional embeddings leads to a significant accuracy boost when predicting the next item. In recent literature, it was reported that a multi-dimensional kernel embedding with temporal contextual kernels to capture users’ diverse behavioural patterns results in a substantial performance improvement. In this study, we further improve the sequential recommender model's robustness and generalization by introducing a mix-attention mechanism with  a layer-wise noise injection (LNI) regularization. We refer to our proposed model as \underline{$ad$}aptive \underline{$r$}obust sequential \underline{$re$}commendation framework $(ADRRec)$, and demonstrate through extensive experiments that our model outperforms existing self-attention architectures.
\end{abstract}
\maketitle



\section{Introduction}
Sequential recommendation plays a pivotal role in enhancing user experience and engagement in various e-commerce and social media platforms. Unlike traditional recommendations that focus solely on static preferences, sequential recommendations leverage the sequential nature of user interactions to provide personalized recommendations over time. By analyzing users' historical behaviors and interactions in chronological order, it is easy to effectively capture temporal dynamics, user preferences, and evolving interests. Sequential recommendation algorithms employ techniques such as recurrent neural networks, convolutional neural networks, and self-attention mechanisms to model sequential patterns and predict users' future actions \cite{hidasi2016session, kang2018self}. 
Recently, several variants of attention-based sequential recommendation models have emerged to enhance the performance of item ID-based models (i.e. SASRec) \cite{kang2018self}. For example, BERT4Rec \cite{sun2019bert4rec} adapts bidirectional self-attention from BERT via masked language modeling. TiSASRec \cite{li2020time} models user interaction sequences with time intervals, and MEANTIME \cite{cho2020meantime} incorporates a mixture of attention mechanisms with multi-temporal embeddings, etc. While these models demonstrate improved performance, they often focus on specific aspects of user interaction sequence modeling, such as time-based or context-based features, attention architecture design, or robustness through techniques like data augmentation or denoising \cite{qiu2022contrastive, chen2022denoising}, etc.  To capture the complexity of the user behavior and provide highly personalized recommendations, it is important to encode each user's unique behavior and apply customized attention mechanisms to learn short-term and long-term user interest as well.\\
In this work, we aim to capture unique sequential patterns of user behavior and prioritize the robustness and generalization of  model as well. Our key contributions are: 
(1) Our proposed method incorporates multi-dimensional kernels for item representation and user behaviors into the attention mechanism; 
(2) Instead of using traditional multi-head attention, we utilized absolute and relative mixture attention mechanism to learn unique patterns from different user behavior; 
(3) NIR (noise injection regularization)  is applied as a layer-wise regularizer to enhance the robustness and generalization; 
(4) The experiments conducted on four popular 
 datasets show our proposed model outperforms the baseline recommenders.

\section{Preliminaries}
In this section, we provide preliminaries and related work.

\textbf{Mix-Attention Mechanism:} 
The multi-head attention plays a crucial role in sequential recommendations to effectively capture long-range dependencies and dynamic relationships within user interaction sequences \cite{kang2018self}. Mix-Attention mechanism proposed in MEANTIME \cite{cho2020meantime} is an extension of the classical multi-head attention mechanism \cite{kang2018self}. Instead of keeping each head processing the split input embedding from the global input matrices, mix-head attention process information of the queries $\mathbf{Q}$, keys $\mathbf{K}$, and values $\mathbf{V}$ for each head from different embedding input scheme (globally or locally). 

\begin{figure*}[!t]
\centering
\includegraphics[width=1\textwidth]{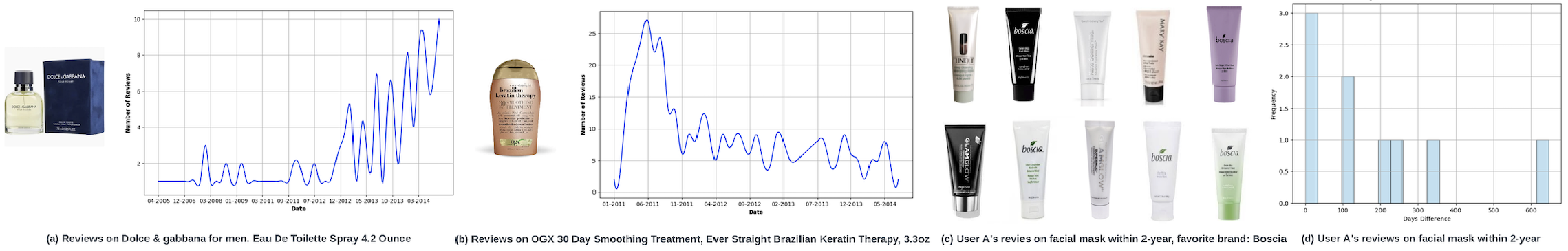}
\caption{
(a) Review trend for a perfume product, showing an increase in reviews from 2005 to 2014 with growing popularity. 
(b) Review trend for 
another product, showing a decline in reviews from 2005 to 2014, possibly indicating reduced consumer interest. 
(c) User's interest in face masks. 
(d) User's review distribution reveals consistent purchasing behavior for facial masks every 3 months, mainly choosing the \textit{Boscia} brand.
(Data Source: Amazon Beauty Review \cite{kang2018self})
}
  \label{fig:abs}
\end{figure*}

\textbf{Multi-dimensional Embeddings:} 
In sequential recommendation, effectively modeling user interactions involves using absolute or relative time-series kernels, absolute or relative positional kernels \cite{rahmani2023incorporating, cho2020meantime, he2021locker}. For instance, absolute time stamps, such as day-of-week and seasonality embeddings, identify patterns in user behavior (Figure \ref{fig:abs}a and b). Additionally, embeddings of relative time intervals reveal how preferences change over different timescales (Figure \ref{fig:abs}c). Furthermore, relative positional-related embedding within a user sequence capture relative positions and transitions, aiding in understanding the progression of user interests. 



\textbf{Robust Regularizer and Exploration:}  
To enhance robustness and generalization in transformer related tasks, some works have focused on designing lightweight transformer architectures\cite{ guo2019star}. 
Others involve using learnable attention distributions \cite{correia2019adaptively, chen2022denoising}.
One more line of approach is adopting noise injection regularization (NIR) schemes, for instance, layer-wise noise stability regularization (LNSR) \cite{hua2023improving} is introduced in an unsupervised manner and provides performance benefits.

\section{Our Learning Framework}
In this section, we introduce our adaptive robust sequential recommendation $(ADRRec)$ model. 
\subsection{Problem Formulation}
Let $U$ be a set of users, and $V$ a set of items. For each user $u \in U$, we have a sequence of item Ids $V_u = [v_{u1}, \ldots, v_{uk}, \ldots, v_{|V_u|}]$ that the user previously interacted with in chronological order and the corresponding time sequence of the interaction $T_u = [t_{u1}, \ldots, t_{uk}, \ldots, t_{|V_u|}]$ that stores the absolute timestamp values. Our goal is to predict the next item $v_{unext}$ that the user $u$ is likely to interact with at the target timestamp $t_{unext}$ based on the given history $(V_u, T_u)$.
\subsection{Input Representation}
Here we describe various types of embeddings leveraged for modeling sequential user behavior.

\textbf{Absolute Time Embedding:} There are two different approaches for computing absolute time embedding, one is embedding-based, and the other one is projection-based \cite{xu2019self, rahmani2023incorporating}. For embedding-based, each timestamp \( t\) is decomposed into multiple components and each component \( t_i (i\in(1,..,k))\) representing different time unit. For instance, $t_i$ can represent
 one of these units: year, month, day, or minute etc. Each \( t_i\) is employed by a learnable embedding $\mathbb{E}_{t_i} \in \mathbb{R}^d $. The multi-dimensional embedding can be given by:
$
\phi(t) = Concat [w_1\mathbb{E}_{t_1} +b_1 , w_2\mathbb{E}_{t_2}+b_2, \ldots, w_i\mathbb{E}_{t_i}+b_i]
$
, where $\{w_i\}_{i=1}^d$ and $\{b_i\}_{i=1}^d$ are learnable parameters. 
For projection-based embedding, it leverages   the translation-invariant time kernel.   The global continuous time is computed by subtracting the
minimum time-stamp and is described by:
$\phi(t)=\mathcal{F}(w_it+b_i)$,
where $\mathcal{F}$ is a periodic activation function, can be referred to $sinusoid$ function and $\{w_i\}_{i=1}^d$ and $\{b_i\}_{i=1}^d$ are learnable parameters. 

\textbf{Relative Time Embedding} 
Relative time embeddings encode the relationship between each interaction pair in the sequence by utilizing temporal difference information. Given a matrix of temporal differences \( D \in \mathbb{R}^{N \times N }\) defined as \( d_{ij} = (t_i - t_j)/\tau \), where \( \tau \) is an adjustable unit time difference and $i\in N, j\in N$ in the given sequence with length N. The encoding functions on \( D \) are similar to \cite{cho2020meantime}. We use three types of embeddings: \(\phi(d_{ij})=\mathcal{F}(w_i{d_{ij}}+b_i)\) is used to learn periodic occurrences, \(\phi(d_{ij})={\text{Exp}(d_{ij}/freq_h)} \) learns the pattern with an downtrend quickly, and \( \phi(d_{ij}) = {\text{Log}(d_{ij}/freq_h)} \) learns the pattern with an uptrend gradually, where  $freq_h$ is the adjustable parameter.

\textbf{Absolute Positional Embedding} Similar to the absolute time embedding, we use two types of absolute positional embeddings: 
(1) the fixed positional encoding $\phi(p)$, where $p$ is the position index , commonly used for transformer \cite{vaswani2017attention, devlin2018bert}; and (2) the learnable positional embedding same as SASRec  \cite{kang2018self}.

\textbf{Relative Distance Embedding}
To mitigate self-attention modules failing to capture short-term user dynamics,  multiple works \cite{he2021locker, shi2023rtisr} apply gaussian prior or learnable weight to correct the importance of items aligning to the current central item. The positional distance matrix \( D \in \mathbb{R}^{N \times N} \) defined as \( d_{ij} = (p_i - p_j) \) and weight \(G\) is denoted as 
 \(g_{ij} = \exp\left( -{(d_{ij} - \mu)^2}/{2\sigma^2} \right)\) and element-wise multiply to attention score matrix: $G \circ \left({\text{Q} \cdot \text{K}^T}/{\sqrt{d_k}}\right)$, where $p_i, p_j \in N$ are the position indices in the given sequence with length N, $\mu$ and $\sigma$ are used for initialization or the weight G can be learnable. 

\subsection{Attention Mechanism}

\textbf{Mix-attention:} We adopt the mix-attention architecture 
as our backbone which is composed of absolute attention and relative attention. We employ multiple absolute and relative kernel embeddings to capture users' diverse patterns.
The absolute embedding attention head is described as: $\text{head}_{\text{a}} = \text{Attention}(\mathbf{Q} + \mathbf{Q^a}, \mathbf{K} + \mathbf{K^a}, \mathbf{V})$,  
where \(\mathbf{Q}\) and \(\mathbf{Q^a}\) are the query matrices composed of the common embedding (e.g., ItemID embedding \(\mathbf{Q}\)) and one or more different absolute embeddings (e.g., \(\mathbf{Q^a}\)), respectively; \(\mathbf{K}\) and \(\mathbf{K^a}\) are the corresponding key matrices; \(\mathbf{V}\) is the value matrix
and the relative embedding attention head can be described as: $\text{head}_{\text{r}} = \text{Attention}(\mathbf{Q}, \mathbf{K}, \mathbf{V}) + \text{Attention}(\mathbf{Q} +\mathbf{b}^r , \mathbf{K^r}, \mathbf{V})$
, where \(\mathbf{K}^r\) represents the relative encoding key matrix; $\mathbf{b}^r $ represents the learnable bias vector added to the queries.

\textbf{Stacking Layer and Point-Wise Feed-Forward Network:} Our model operates similar to \cite{kang2018self}. We apply Position-wise Feed Forward Network (FFN) and stack \(L\) sublayers with residual connection including LayNorm() and Dropout() as well.

\subsection{Robust Input and Output }
Inspired by NIR \cite{hua2021noise}, noise is injected into the transformer layers during training, and explicit layer regularization is applied. Specifically, given an input point \( x \), a perturbed input \( \bar{x} \) is generated by adding random noise \( \epsilon \) with a small magnitude to \( x \). We enhance the strategy by creating a parameterized neural network for learnable injected noise. Considering a linear neural network with \( m \) inputs and \( n \) outputs, denoted by $\hat{x} = w \cdot x + b$,
where \( x \in \mathcal{R}^m \) is the layer input, \( w \in \mathcal{R}^{m \times n} \) is the weight matrix, and \( b \in \mathcal{R}^n \) is the noise. The corresponding neural layer with respect to the input \( x \) can be represented as: $\hat{x} = (\mu_w + \sigma_w \odot \epsilon_w) \odot x + \mu_b + \sigma_b \odot \epsilon_b$
The parameters $\mu_w \in \mathbb{R}^{m \times n}$, $\mu_b \in \mathbb{R}^n$, $\sigma_w \in \mathbb{R}^{m \times n}$, and $\sigma_b \in \mathbb{R}^n$ are learnable, $\epsilon_w \in \mathbb{R}^{m \times n}$ and $\epsilon_b \in \mathbb{R}^m$ are standard gaussian noise random variables, where \( \zeta \) =\( (\mu, \sigma) \) is the set of learnable vectors. We apply layerwise noise stability regularizer (LNSR) on the training data set to enhance robustness and generalization following: $\mathcal{\hat{R}}(\theta) = \mathbb{E}_{\epsilon} \| f(x + \epsilon) - f(x) \|^2 = \sum_{l=1}^{L} \lambda_{i,j} \| f_{i,j}(\mathbf{x} + \boldsymbol{\varepsilon}) - f_{i,j}(\mathbf{x}) \|^2$. 
, where $j$ is the layer where noise is injected, $i,j$ are the layer index from $1$ to $L$ (the total number of layers), $\lambda_{i,j}$ are the regularization weights for each layer, $\mathbf{x}$ is the input and $\boldsymbol{\varepsilon}$ is the injected noise vector.
For \(\mathcal{\hat{R}}\), by using first-order and second-order Taylor expression to represent  $f(\mathbf{x} + \boldsymbol{\varepsilon})$, it has: 
$ \mathcal{\hat{R}}(\theta)=\sum \{\Omega_J(f)+ \Omega_H(f)\}$, where $\Omega_J(f)$ and $\Omega_H(f)$ refer to the Jacobian and Hessian of $f$ with respect to the input $x$.   
The regularizer guarantees to be positive by involving the sum of squares of the first-order and second-order derivatives \cite{grandvalet1997noise}. 
\subsection{Learning Objective}
Our model $(ADRRec)$ training is composed of two components: (1) absolute and relative pattern training to learn user long-term and short-term preferences and (2) stability regularizer for generalization. The form of the cost function can be represented as: $\theta^* = \arg \min_{\theta} \mathbb{E} \left[ \mathcal{L}(f(x; \theta), y) + \lambda \hat{R}(\theta) \right]$
, where  $\lambda \in [0,1]$ and $\mathcal{L}$ is the loss function measuring the discrepancy between the network's prediction $\tilde{f}(\mathbf{x}; \theta)$ and the true label $\mathbf{y}$.  $\hat{R}(\theta) $ is the regularizer.
\begin{table}[!t]\footnotesize
\centering
\caption{Datasets Statistics}
\label{tab:dataset-stats}  
\begin{tabularx}{0.45\textwidth}{lXXXXX}
\toprule
Dataset & \#users & \#items  & Avg. length &\#actions \\
\midrule
Amazon Beauty & 52,024 & 57,289 & 8.9 & 0.4M \\
Amazon Games & 31,013 & 23,715  & 6.88 & 0.5M \\
MovieLens-1M & 6,040 & 3,416  & 165.56& 0.99M \\
MovieLens-20M & 138,475 & 18,166  & 144.44& 19.7M \\
\bottomrule
\end{tabularx}
\end{table}
\begin{table}[!t]\footnotesize
\caption{Embedding Notation}
\label{tab:notation}
\centering
\begin{tabular}{@{}ll|ll@{}}
\toprule
\#  & Definition & \#  & Definition \\
\midrule
p  & PositionalEmbedding & b & BochnerTimeEmbedding \\
t  & AbsoluteTimeEmbedding & s  & SinusoidTimeDiffEmbedding \\
e &  ExponentialTimeDiffEmbeddin & l  & Log1pTimeDiffEmbedding \\
r  & RelativeDistanceEmbedding & o  & NoiseRegularizer \\
\bottomrule
\end{tabular}
\end{table}
\begin{table*}[t]
\footnotesize
\centering
\caption{Model Comparison: compared ADRRec with four Baseline Models, best baselines (underline), best values (bold) and  relative improvement over the best baselines.}
\label{tab:overall}
\begin{tabular}{@{}l@{}cccccccccccccccc@{}}
\toprule
\multirow{3}{*}{Models} & \multicolumn{4}{c}{Beauty} & \multicolumn{4}{c}{Game} & \multicolumn{4}{c}{ml-1m} & \multicolumn{4}{c}{ml-20m} \\
\cmidrule(r){2-5} \cmidrule(r){6-9} \cmidrule(r){10-13} \cmidrule(r){14-17}
& NDCG & NDCG & Recall & Recall & NDCG & NDCG & Recall & Recall & NDCG & NDCG & Recall & Recall & NDCG & NDCG & Recall & Recall \\
& @5 & @10 & @5 & @10 & @5 & @10 & @5 & @10 & @5 & @10 & @5 & @10 & @5 & @10 & @5 & @10 \\
SASRec & 0.1333 & 0.1550 & 0.1873 & 0.2547 & 0.2368 & 0.2768 & 0.3416 & 0.4654 & 0.3968 & 0.4419 & 0.5459 & 0.6846 & 0.3733 & 0.4250 & 0.2073 & 0.2747 \\
TISAS & 0.1541 & 0.1752 & 0.2045 & 0.2318 & 0.2718 & 0.3114 & 0.3641 & 0.5068  & 0.4031 & 0.4447 & 0.5516 & 0.6796 & 0.3941 & 0.4552 & 0.2345 & 0.2804 \\
BERT4Rec & 0.1597 & 0.1848 & 0.2223 & 0.3001 & 0.2982 & 0.3445 & 0.4227 & 0.5662 & 0.4181 & 0.4579 & 0.5628 & 0.6857 & 0.4006 & 0.4456 & 0.5447 & 0.6832 \\
MEANTIME & \underline{0.1644} & \underline{0.1899} & \underline{0.2286} & \underline{0.3071} & \underline{0.3208} & \underline{0.3671} & \underline{0.4533} & \underline{0.5945} & \underline{0.4432} & \underline{0.4808} & \underline{0.5920} & \underline{0.7079} & 
\underline{0.4042} & \underline{0.4499} & \underline{0.5508} & \underline{0.6917} \\
ADRRec & \textbf{0.1749} & \textbf{0.2013} & \textbf{0.2423} & \textbf{0.3244} & \textbf{0.3299} & \textbf{0.3729} & \textbf{0.4616} & \textbf{0.5966} & \textbf{0.4610} & \textbf{0.4972} & \textbf{0.6132} & \textbf{0.7274} & 
\textbf{0.4165} & \textbf{0.4608} & \textbf{0.5626} & \textbf{0.6992} \\
\midrule
Improvement & 6.39\% & 6.00\% & 5.99\% & 5.63\% & 2.84\% & 1.58\% & 1.83\% & 0.35\% & 4.02\% & 3.41\% & 3.58\% & 2.75\% & 3.04\% & 2.42\% & 2.14\% & 1.08\% \\
\bottomrule
\end{tabular}
\end{table*}

\begin{algorithm}[h]
    \SetAlgoLined
    \KwIn{Training set $D$, perturbation bound $\delta$, learning rate $\tau$, number of layers $L$, number of training epochs $N$, self-attention $self()$ and model parameters $\theta$, regularization weights for each layer $\{\lambda_{k}, \dots, \lambda_L\}$}
    Initialize $\theta$\;
    \For{$epoch = 1, 2, \dots, N$}{
        \For{minibatch $B \sim D$}{
            $\mathcal{R} \leftarrow 0$\;
            \ForEach{$(x,y) \in B$}{
                Sample noise $\epsilon \sim \mathcal{N}(\mu, \sigma^2)$ from noisy network layer\;
                $\tilde{x} \leftarrow x + \epsilon$\;
                Global and Local multi-dimensional Represent kernel Embedding\;
                $x_e, \tilde{x_e} \leftarrow emb(x), emb(\tilde{x})$\;     
                Perform forward pass given $x_e$ and $\tilde{x_e}$ as inputs\;
                \For{$r = k, k + 1, \dots, L$}{
                    $\mathcal{R} \leftarrow \mathcal{R} + \lambda_{k} \|self_k(x) - self_k(\tilde{x})\|^2$\;
                }
            }
            $g \leftarrow \frac{1}{|B|} \sum_{(x,y) \in B} \nabla_{\theta} [L(f(x;\theta), y) + \mathcal{R}]$\;
            $\theta \leftarrow \theta - \tau g$\;
        }
    }
    \KwOut{$\theta$}
    \caption{Training Robust Adaptive Sequential Recommendation}
\end{algorithm}

\section{Experiments}
In the experiments, we aim to address several key questions: (1) Comparison with Baseline Models: How does $ADRRec$ perform compared to baseline models in terms of accuracy? (2) Impact of Different Components: What is the effect of incorporating different components, such as absolute and relative embeddings, on the model's performance? (3) Robustness Analysis: How robust is $ADRRec$ to variations in the input data? We evaluate our approach on four real-world datasets: MovieLens 1M and 20M, Amazon Beauty and Game \cite{kang2018self}.
\subsection{Comparison with Baseline Models}
To validate the effectiveness of our proposed model, we compare it with popular baselines: SASRec \cite{kang2018self}, BERT4Rec \cite{devlin2018bert}, TISAS \cite{li2020time}, and MEANTIME \cite{cho2020meantime}. These models were selected due to their strong performance in sequential recommendation tasks, as they leverage self-attention mechanisms: SASRec using left-to-right attention, BERT4Rec using bidirectional attention, and TiSASRec accounting for temporal intervals and MEANTIME utilize mixture attention framework—making them well-suited for our comparative analysis. Among the baseline methods, our model $ADRRec$ consistently outperforms the rest baselines (Table ~\ref{tab:overall}).
\subsection{Robustness Analysis}
To evaluate  robustness and generalization of our model, we performed robustness studies by: (1) evaluating the standard deviation of the model results run with 3 random seeds (Table \ref{tab:reg}). The results on mean, std show the stability and robustness. (2) In the test set, randomly masking some continuous portion (10\%, 30\%) in input test sequence (Table \ref{tab:ood}). We observe  it does not degrade generalization performance.
\begin{table}
\footnotesize
\caption{Robustness Evaluation (Mean, Std). Mode: (p-s-l-e)}
\label{tab:reg}
\centering
\begin{tabular}{@{}lcccccccc@{}}
\toprule
\multirow{2}{*}{Metrics} & \multicolumn{2}{c}{Beauty} & \multicolumn{2}{c}{Game} & \multicolumn{2}{c}{ml-1m} & \multicolumn{2}{c}{ml-20m} \\
\cmidrule(r){2-3} \cmidrule(r){4-5} \cmidrule(r){6-7} \cmidrule(r){8-9}
& mean & std & mean & std & mean & std & mean & std \\
\midrule
NDCG@5 & 0.1716 & 0.0009 & 0.266 & 0.0009 & 0.4587 & 0.007 & 0.4156 & 0.0010\\
NDCG@10 & 0.1986 & 0.0007 & 0.3104 & 0.0006 & 0.7335 & 0.006 & 0.4561 & 0.0008 \\
Recall@5 & 0.2388 & 0.0013 & 0.3825 & 0.0004 & 0.8268 & 0.005 & 0.5530 & 0.0004 \\
Recall@10 & 0.3228 & 0.0010 & 0.5186 & 0.0003 & 0.8750 & 0.002 & 0.6964 & 0.0015 \\
\bottomrule
\end{tabular}
\end{table}

\begin{table}[!h]\footnotesize
\caption{Robustness Evaluation (OOD). Mode: (p-s-l-e)}
\label{tab:ood}
\centering
\begin{tabular}{@{}ccccccccc@{}}
\toprule
\multirow{2}{*}{Metrics} & \multicolumn{2}{c}{Beauty} & \multicolumn{2}{c}{Game} & \multicolumn{2}{c}{ml-1m} & \multicolumn{2}{c}{ml-20m} \\
\cmidrule(r){2-3} \cmidrule(r){4-5} \cmidrule(r){6-7} \cmidrule(r){8-9}
& 10\% & 30\% & 10\% & 30\% & 10\% & 30\% & 10\% & 30\% \\
\midrule
NDCG@5 & 0.1513 & 0.1481 & 0.2613 & 0.2613 & 0.4478 & 0.4449 & 0.4071 & 0.4004 \\
\midrule
NDCG@10 & 0.1782 & 0.1748 & 0.3042 & 0.3034 & 0.4847 & 0.4816 & 0.4523 & 0.4456 \\
\midrule
Recall@5 & 0.2106 & 0.2065 & 0.3736 & 0.3723 & 0.5948 & 0.5924 & 0.5527 & 0.5452 \\
\midrule
Recall@10 & 0.2943 & 0.2895 & 0.5040 & 0.5053 & 0.7079 & 0.7057 & 0.6923 & 0.6846 \\
\bottomrule
\end{tabular}
\end{table}

\begin{table}[!h]\footnotesize
\caption{ ADRRec w/wo Kernel Embeddings and NIR on combination of embedding components }
\label{tab:abcomp}
\centering
\begin{tabular}{@{}l@{\hspace{1.4mm}}l@{\hspace{1.4mm}}l@{\hspace{1.4mm}}c@{\hspace{1.4mm}}c@{\hspace{2mm}}l@{\hspace{1.4mm}}l@{\hspace{1.4mm}}l@{\hspace{1.4mm}}c@{\hspace{1.4mm}}c@{}}
\toprule
& \multicolumn{4}{c}{Beauty} & & \multicolumn{4}{c@{}}{ml-1m} \\
\cmidrule(r){1-5} \cmidrule(l){6-10}
\multirow{2}{*}{Mode} & \multicolumn{2}{c}{NDCG} & \multicolumn{2}{c}{Recall} & \multirow{2}{*}{Mode} & \multicolumn{2}{c}{NDCG} & \multicolumn{2}{c@{}}{Recall} \\
& @5 & @10 & @5 & @10 & & @5 & @10 & @5 & @10 \\
\midrule
p-b-l-e-o & 0.1713 & 0.1985 & 0.2392 & 0.3269 & p-b-l-e-o & 0.4422 & 0.4807 & 0.5888 & 0.7065 \\
p-b-l-e & 0.1702 & 0.1985 & 0.2387 & 0.3229 & p-b-l-e & 0.4377 & 0.4754 & 0.5859 & 0.7017 \\
p-b-s-l-o & 0.1695 & 0.1964 & 0.2358 & 0.3194 & p-b-s-l-o & 0.4517 & 0.4874 & 0.6002 & 0.7105 \\
p-b-s-l-r-o & 0.1749 &0.2013 & 0.2423 & 0.3244 & p-b-s-l-r-o & 0.4610 & 0.4972 & 0.6132 & 0.7242 \\
\bottomrule
\end{tabular}
\end{table}
\subsection{Ablation Studies}
To evaluate the impact of different components, we used Beauty and ML-1M datasets to compare the performance: (1) w vs. w/o absolute/relative kernel embedding, and (2) w vs. w/o noise regularizer. Table ~\ref{tab:abcomp} shows the comparison of different combinations of  kernel embedding plus noise injection regularizer. The results show noise injection regularizer have effectiveness in both datasets. 

\section{Conclusion}
In this paper we presented $ADRRec$, an adaptive and robust sequential recommendation model, which uses  multi-dimensional kernel encoding and mix-attention mechanism to learn each unique user behavior. We apply layer-wise noise injection regularization to enhance robustness and generalization. Experiments on four classical datasets show that our model outperforms the baselines in sequential recommendation. 


\bibliographystyle{ACM-Reference-Format}
\bibliography{main-short}


\end{document}